# A six degree of freedom nanomanipulator design based on carbon nanotube bundles

**Vasilii I Artyukhov**
Institute of Biochemical Physics, Russian Academy of Sciences
Kosygin st. 4, Moscow, 119334 Russia

E-mail: artyukhov@sky.chph.ras.ru

**Abstract.** Scanning probe imaging and manipulation of matter is of crucial importance for nanoscale science and technology. However, its resolution and ability to manipulate matter at the atomic scale is limited by rather poor control over the fine structure of the probe. In the present communication, a strategy is proposed to construct a molecular nanomanipulator from ultrathin single-walled carbon nanotubes. Covalent modification of a nanotube cap at predetermined atomic sites makes the nanotube act as a support for a functional “tool-tip” molecule. Then, a small bundle of nanotubes (3 or 4) with aligned ends can act as an extremely high aspect ratio parallel nanomanipulator for a suspended molecule, where protraction or retraction of individual nanotubes results in controlled tilting of the tool-tip in two dimensions. Together with the usual SPM three degrees of freedom and augmented with rotation of the system as a whole, the design offers six degrees of freedom for imaging and manipulation of matter with precision and freedom so much needed for advanced nanotechnology. A similar design might be possible to implement with other high-aspect ratio nanostructures, such as oxide nanowires.

## 1. Introduction

Scanning probe imaging and manipulation of matter crucially relies on the quality of tips, specifically, on their aspect ratio, which determines the spatial resolution, and their wearing behavior, which determines the reliability and lifetime of a probe. From this point of view, carbon nanotubes (particularly, single-wall nanotubes [1, 2]) are the ideal candidate due to their small and uniform diameter and extremely strong graphitic bonds between constituent atoms. Moreover, carbon nanotubes can be either semiconducting or metallic, and the latter is crucial for scanning tunneling microscopy and related atomic manipulation techniques.

Immediately after the first demonstration of carbon nanotube usage as scanning microscope probes in 1996 [3], the field literally exploded with various applications and modifications of the technology. Lithography with carbon nanotubes was demonstrated in 1998 [4]. Manipulation of nanoscale objects by multiple-nanotube devices was achieved as early as 1999 [5], promptly followed by fabrication of nanostructures via controlled deposition of specific length segments of the nanotube tips [6]. Covalently functionalized carbon nanotube tips were used to demonstrate scanning chemical force microscopy [7] and chemically sensitive tunneling microscopy [8]. Magnetic force microscopy can be performed using carbon nanotubes functionalized with magnetic metal nanoclusters [9]. Pristine and chemically modified carbon nanotube tips have earned their place as an important tool for scanning probe microscopy in both physical [10] and biological science [11]. However, their application is still limited by various factors: on one hand, the difficulties of positioning and attachment of nanotubes to tips complicate the process of tip fabrication and attachment of the nanotube to the cantilever is often not too reliable, which makes conventional silicon probe technology more practical for most current applications where the extreme aspect ratio of the probe offered by nanotubes is not absolutely essential; on the other hand, there exist fundamental issues such as the fact that the exact type of grown nanotubes is hard to control, and the spatial unpredictability of covalent functionalization, meaning that it is impossible to tell in advance the exact location of the functional group or molecule. In the following, a strategy is proposed to circumvent the latter problem, which ultimately results in the proposal of a novel family of nanoscale parallel manipulators with two extra degrees of freedom compared to the traditional scanning probe technology, hoping that projected gains from such tools can outweigh the associated short-term practical difficulties and stimulate further research effort to mitigate these problems.

## 2. Nanomanipulator design

### *2.1. Functionalization of nanotubes at predetermined sites: the isolated pentagon rule*

It is well known that in order to form a closed surface, an $sp^2$ carbon nanostructure must contain 12 pentagons that create positive surface curvature; therefore, a hemispherical nanotube cap contains 6 pentagons. The angle of an equilateral pentagon is 108°, which deviates strongly from the optimal value of 120° for $sp^2$ hybridization of carbon, and is in fact closer to the 109.5° value that is typical for tetrahedral $sp^3$ hybridization. Therefore, an atom that belongs to two or three pentagons at the same time is "forced" into an $sp^3$-like state by this geometrical strain, which creates effectively unpaired electron density at the atom. Such an atom would be prone to become engaged in addition reactions so that its chemical environment builds up to an $sp^3$ tetrahedron.

This result is well-known as the "isolated pentagon rule" (IPR). It explains why the smallest stable fullerene is $C_{60}$, and it only exists in the truncated-icosahedron $C_{60}$ isomeric form that obeys the IPR (i.e., all pentagons are separated by hexagons). Smaller fullerenes, such as the $C_{20}$ dodecahedron containing only pentagons, demonstrate poor stability; at the same time, the dodecahedrane molecule ($C_{20}H_{20}$) is stable.

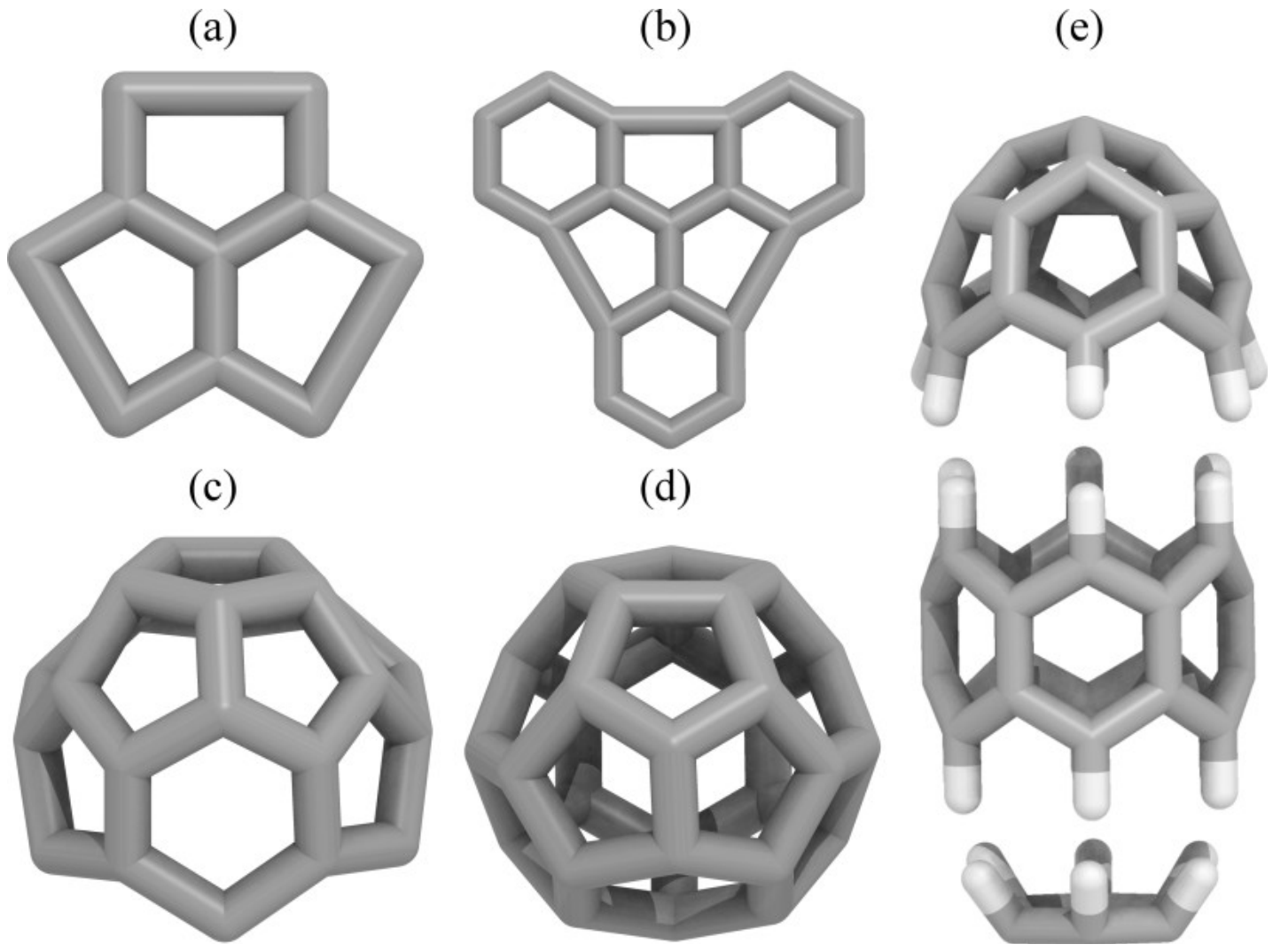

**Figure 1.** (6,0) nanotube caps made from the $C_{28}$ fullerene: (a-d) construction of the fullerene starting from three pentagons sharing a common atom; (e) cutting the fullerene into two caps and insertion of a (6,0) nanotube segment inbetween. The upper cap contains the three-pentagon cluster.

Since $C_{60}$ is the smallest fullerene that has isomers obeying the IPR, any nanotube with a diameter smaller than 0.68 nm (the diameter of $C_{60}$) will inevitably contain edge- or vertex-sharing pentagons in its cap. These sites will be the most chemically active, making them prime candidates for chemical functionalization. We can further imagine a situation where a nanotube cap contains a *unique* most preferred spot for covalent functionalization; for this, we need to recall the structure of the $C_{28}$ fullerene, which contains four clusters of three pentagons sharing a vertex ($T_d$ point symmetry group). The clusters are linked by common pentagon edges. The process of constructing a $C_{28}$ fullerene is shown in four steps starting from our cluster of interest in figure 1 (a–d). The final structure can be sliced in half so that two (6,0) nanotube caps are formed; this is shown in figure 1 (e), where a segment of the nanotube has been inserted between the halves. The upper fragment has a threefold rotational axis ($C_{3v}$ group), while the symmetry of the lower is sixfold ($C_{6v}$).

Since no experimental data exist on the geometry of caps (or whether at all caps can be stable) in such small-diameter nanotubes, the question of which of the two possible cap structures is the correct one has to be addressed using quantum chemical calculations. The relative stability of the two different cap structures can be roughly judged from the deformation energies of corresponding fullerenes that are made by joining two caps of each type. The upper cap structure corresponds to one of the isomers of $C_{44}$, the lower corresponds to $C_{36}$. Calculations (see table 1) show that our cap structure is preferred for the (6,0) nanotube since it minimizes the energy per atom: it has one atom in the most unfavorable position possible—shared by three pentagons—but this is overcompensated by there being fewer edge-sharing pentagons (3 common edges instead of 6). It should also be noted that the "bottom" cap can be converted into a "top" cap by adding 4 carbon atoms. The data in table 1 demonstrate that the energies-per-atom of such caps come quite near to that of the (stable) $C_{60}$ fullerene; chemical functionalization of the tip atom should be expected to substantially stabilize the $C_{3v}$ cap structure even further.

**Table 1.** Relative energies of possible (6,0) nanotube caps.

| Fullerene | Binding energy per atom (kJ/mol)[a] |
|---|---|
| $C_{28}$ (figure 1d) | –671.3 |
| $C_{36}$ (two $C_{6v}$ caps [12]) | –691.3 |
| $C_{40}$ (one cap of each type) | –691.4 |
| $C_{44}$ (two $C_{3v}$ caps) | –695.8 |
| $C_{60}$ | –720.0 |

[a] Calculations were performed using the PBE density functional [13] with an optimized triple-zeta Gaussian basis set.

Thus, the (6,0) carbon nanotube is the perfect candidate for use as a chemically functionalized probe since its cap contains a single site, referred to as the "tip atom" in the following, that is especially susceptible to chemical functionalization. The diameter of this nanotube is 0.47 nm, suggesting a very favorable aspect ratio. Incidentally, this tube is metallic, meaning that the operation of the functional group at the tip could be controlled by applied voltage. Another point to note with regard to the (6,0) nanotube is its good lattice match with the (111) diamond surface [14], suggesting the use of diamond cantilevers to support the nanotubes. As for other nanotube types, the complete inventory of carbon nanotube cap structures (up to a diameter of 3 nm) available in the literature [15] should facilitate the search for other systems that offer at least partial site-specificity of chemical functionalization and hence could be used in other designs resembling the present.

To conclude the section dedicated to the isolated pentagon rule, it should be noted that substitution of carbon with nitrogen, in a sense, reverses the IPR. The nitrogen atom has an extra electron compared to carbon, and therefore, instead of an effectively unpaired electron, a nitrogen atom shared by three pentagons presents a lone pair. Such a configuration is, in fact, preferential for N-substituted fullerenes [16]. This means that in a capped (6,0) nanotube, a nitrogen atom would predominantly occupy the position at the very tip. While this would actually prevent covalent functionalization of the nanotube at the tip, the nitrogen atom could form a dative bond with, e.g., a boron atom, opening up broad possibilities for reversible assembly and disassembly of complex architectures with boron-substituted carbon nanostructures or with boron nitride structures. Finally, similar arguments should generally hold for boron substitution of nanotubes, i.e., boron atoms should also predominantly occupy the tip position in a capped (6,0) nanotube.

### *2.2. Parallel nanomanipulators based on nanotube bundles*

While site-specific functionalization of carbon nanotubes may be interesting for imaging and manipulation of matter at the scale of individual atoms, the above strategy becomes effectively useless if deterministic manipulation of multi-atom molecules is required: free rotation of the grafted molecule about the $sp^3$ bond would render null all the efforts to lock it into the desired location. However, this problem can be solved by simultaneously attaching the molecule to three (or more) nanotubes forming a bundle.

As an illustration, an adamantane molecule attached to a bundle of three capped (6,0) nanotubes is shown in figure 2 (a). This molecule was chosen for its tetrahedral shape giving it three natural linking sites and a fourth that remains free for subsequent modification. The relatively small size of the molecule means that intermediary linkers are required. A number of possible linking configurations using carbon, silicon, germanium, oxygen and sulfur atoms were tested; from these, alkane chains with three carbon atoms (i.e., propane) appear to be most suitable, long enough to make up for the small size of the molecule compared to the gap between nanotube tips, but not too long so that the molecule is kept in place tightly.

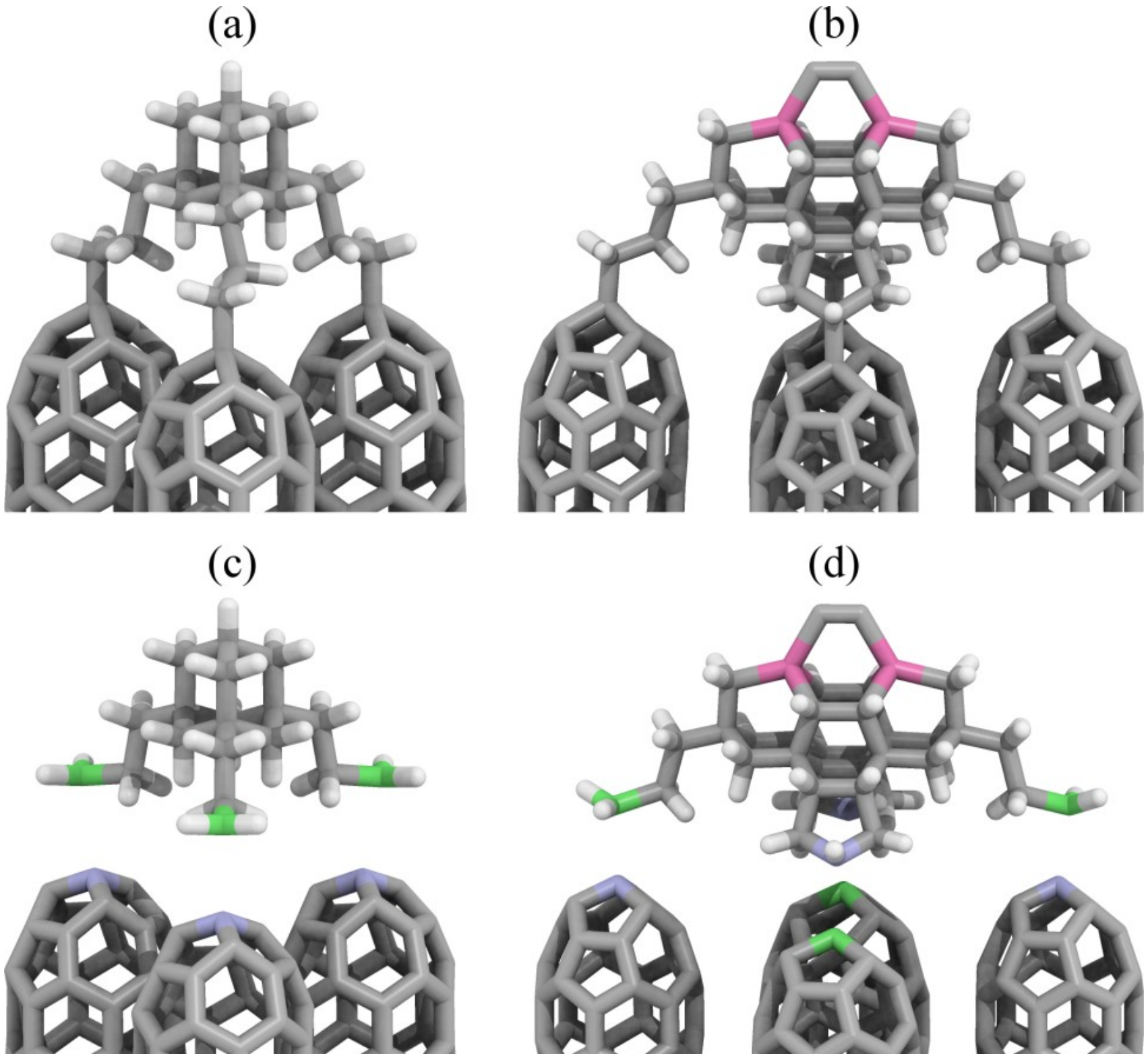

**Figure 2.** Carbon nanotube bundles covalently grafted with (a) an adamantane molecule and with (b) a carbon dimer deposition tool based on two face-joined adamantane molecules. Dative attachment is also possible (c, d). Boron and nitrogen atoms are shown in green and blue, respectively; violet represents germanium.

It is also immediately seen that in such a configuration, relative protraction or retraction of individual nanotubes can tilt the molecule with two angular degrees of freedom. The nanotubes in the bundle are kept together by attractive dispersion forces, but the relative sliding should be easy. Therefore, three-site grafting of the functional molecule, in fact, converts a simple probe into a full-fledged parallel nanomanipulator (see figure 3).

Another reason why adamantane has been chosen as the molecule of interest in the present study is the recent proposal of a minimal toolset for positionally controlled diamond mechanosynthesis by Freitas and Merkle [17]. Among the 9 proposed functional molecules ("tooltips"), 7 represent adamantane molecules with appropriate substitutions of either hydrogen or carbon at one vertex of the molecule, and yet another one is basically a combination of two such tooltips. The final member of the set is the dimer placement tool (DimerP) [18, 19] based on two face-joined Ge-substituted adamantane molecules (this corresponds to a diamond crystal twin boundary) shown in figure 2b suspended on four (6,0) nanotubes. In this case, the molecule connects to the inner nanotubes via cyclopentane rings; furthering the twin boundary analogy, each ring can be viewed as dual propane chains sharing one end and joined at the other.

Examples of the same two molecules linked via dative B–N bonds are also provided in figure 2 (c, d). Notice that while adamantane is supported on three nitrogen-substituted nanotube caps, the mirror symmetry of the DimerP molecule would cause problems if boron substitution were to be used in the cyclopentane rings; however, using two boron-substituted nanotubes solves the problem. In these systems, the functional molecules are bound less strongly than in the covalent case. On the other hand, tooltips can be changed relatively easily. One can immediately see that it is straightforward to use various combinations of boron/nitrogen substitutions to design complementary tools that bind selectively to their intended counterparts, which might turn out useful in future complex architectures.

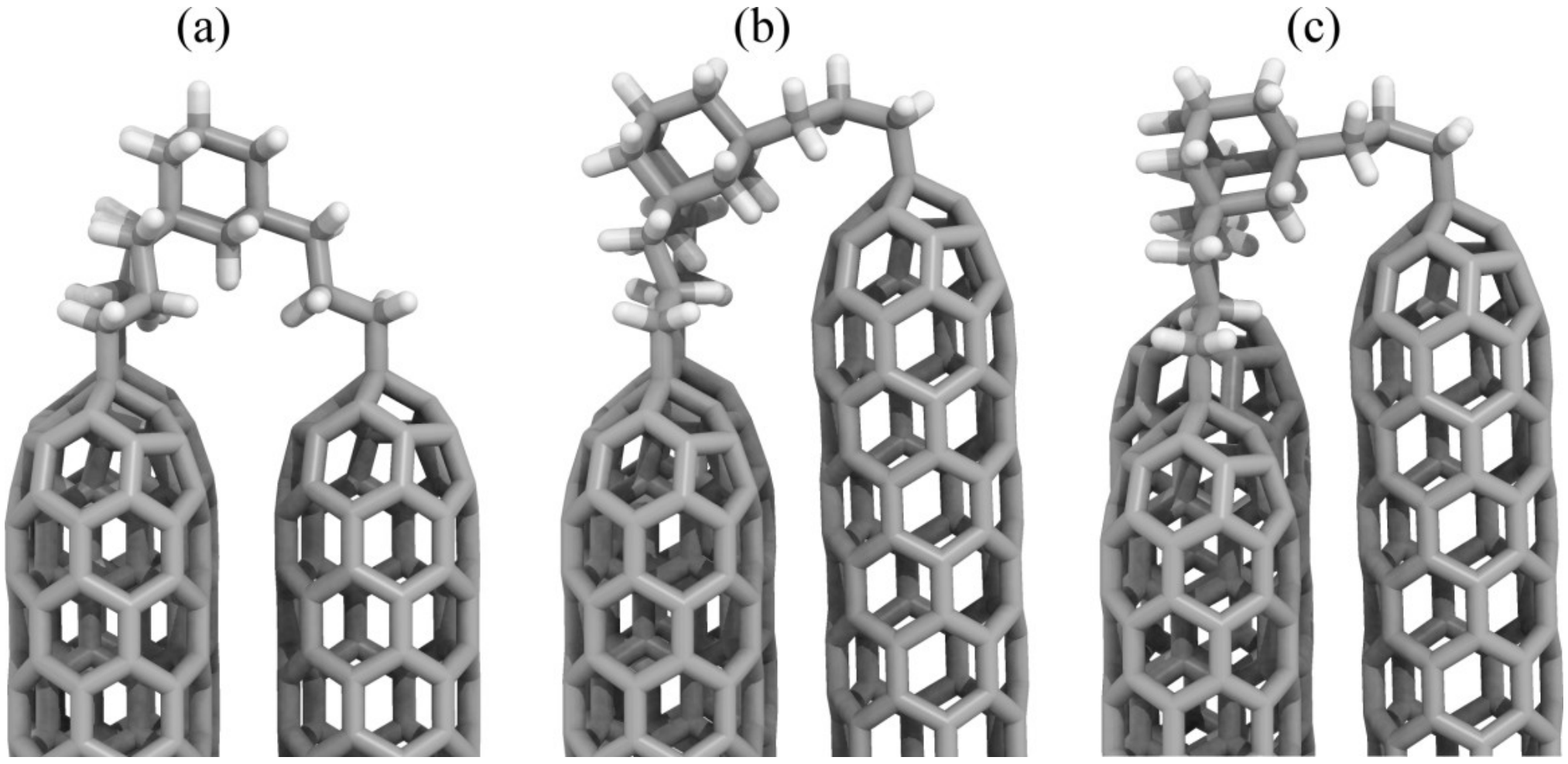

**Figure 3.** Angular flexibility of the manipulator: (a) starting position; (b) one nanotube protracted by ca. 0.5 nm; (c) another nanotube protracted by ca. 0.2 nm. The corresponding strain energies are listed in table 2.

Although a detailed technical assessment of the performance of the manipulators in terms of range of motion, positional and angular uncertainty, etc. is beyond the scope of the present communication, some estimate of the amount of strain present in the systems is nevertheless needed to check if these structures could at all exist. Strain energies were (very roughly) estimated using the MM2 molecular mechanics force field [20], and the results are summarized in table 2. It can be seen that, although a certain amount of strain is present in all structures, it is insufficient to cause bond rupture, especially considering the fact that it is distributed over 3 (for adamantane) or 4 (for DimerP) links. This also means that bond configurations in the structures are not too unusual, and the use of molecular mechanics (avoiding expensive quantum-chemical calculations) is justified in this case. In summary, these results show that the designs in figure 2 may be feasible, at least from the thermodynamical point of view, and should work as expected (figure 3).

**Table 2.** Strain energy estimates.

| Structure | Energy (kJ/mol)[a] |
|---|---|
| Adamantane (untilted) | 232 |
| Adamantane (tilted, figure 3b) | 280 |
| Adamantane (tilted, figure 3c) | 260 |
| DimerP (untilted) | 307 |
| Typical alkane C-C bond strength | ~350 |

[a] Energies are listed with respect to nonfunctionalized nanotube bundles and molecules with pre-attached linkers.

## 3. Discussion

*3.1. Implementation pathways*

Before the implications of the above designs can be discussed, possible strategies of fabricating the proposed structures have to be reviewed. This includes synthesizing the required components and assembling them into a working system.

As of present, carbon nanotube probes are typically grown *in situ* on SPM tips using some variation of chemical vapor deposition process [21], with the possibility of even wafer-scale fabrication [22]. However, CVD-grown nanotubes typically have diameters > 1 nm, which is too large for our purposes, and their type is hard to control precisely. On the other hand, ultrathin SWCNTs down to ~0.4 nm diameter can be selectively grown inside zeolite pores [23–26], or inside larger diameter CNTs [27] with the possibility of controlling the type of as-grown nanotube by the choice of catalyst type and external conditions [28]. The inner tube could subsequently be extracted from the resulting double-wall nanotube by mechanical means (so-called "sword-in-sheath" failure of the outer wall) [29] or, for example, using electrical current heating [30]. Even if the nanotubes are grown uncapped, it should nevertheless be possible to close their ends; on-demand capping of carbon nanotubes has previously been demonstrated, at least, for multiwall carbon nanotubes [31].

Given all the difficulties of fabrication and processing of ultrathin carbon nanotubes, it might be desirable to use nanocones [32] or conically-terminated multiwall nanotubes, since these structures can have very sharp tips [33] with clusters of pentagons. Although chemical modification of nanocones is much less explored compared to nanotubes, quantum chemical calculations [34] suggest that functionalization of nanocones should occur predominantly at the tip, offering at least some spatial control over functionalization. Finally, it should be noted that perfect control over the functionalization site is not an absolute necessity: techniques such as field emission measurements with a second movable probe [35] could in principle be utilized to determine functional group position after the functionalization has been carried out, thus enabling the use of other carbon nanostructures besides the (6,0) nanotube.

Individual as-grown carbon nanotubes will then have to be transferred onto separate actuators, and their free ends joined together to form a self-supporting bundle. Although in principle, just one degree of freedom per actuator should be sufficient—provided that the tubes are long enough (or the actuators close enough) so that they can be joined using an additional 3-dof manipulator,—assembly would be most easily done if all actuators had three degrees of freedom. This suggests the use of a three-probe scanning microscope design for the first demonstrations of the devices; it should then be noted that in the present case, steric hindrance constraints that plague conventional multiprobe instrument design are somewhat relaxed, because the probes only need to approach each other to a certain distance (determined by the length of the nanotubes); nor need they be parallel or coplanar, providing additional design flexibility to reduce steric congestion.

After the bundle has been formed, for example, DNA hairpins [36] could be used to make a "knot" clipping the bundle together and allowing individual nanotubes to be routed to their independent actuators, although it is quite probable that it would be sufficient to simply rely on the mutual attraction of carbon nanotubes. Any excess length of the nanotubes could be trimmed *in situ* with, e.g., an electron beam [37].

Covalent functionalization of carbon nanotubes is a well-established procedure, and, as long as there exists a preferred spot of functionalization on the nanotube, no insurmountable obstacles are to be expected from this side. Similarly, organic synthesis methods are more than capable of producing molecules with appropriate linkers attached, as long as a desired functional molecule has been chosen. The possibility of successful synthesis of particular tooltip molecules discussed above has already been addressed in the corresponding references.

Given the freedom to choose the functional groups on both sides—nanotube tips and the molecules—it appears that the rest (putting the functionalized molecule on a pre-assembled functionalized nanotube bundle) is also within the reach of scanning probe manipulation technologies.

*3.2. Design variations*

In the above, only three degrees of freedom of the systems have been explicitly considered, namely, those associated with the vertical (along the bundle axis) translations of individual nanotubes. These correspond to the vertical translation of the system as a whole and two Euler angles. Three more degrees of freedom that have to be introduced ‘externally’ are two horizontal translations and the rotation about the bundle axis. These could be split between the manipulator and the substrate, hopefully providing some room to simplify manipulator design. Moreover, one nanotube could be kept fixed to save some complexity on its actuator: the nanotube actuators appear to be the most troublesome spot of the whole system since they have to be made both very small and very precise, and transferring this degree of freedom to the substrate or to manipulator suspension could be very helpful from the engineering point of view. Alternatively, nanotube ends could be statically anchored on a single platform having two “tilt” degrees of freedom. Overall, given the recent progress in the design of complex and precise nanomanipulators (including multistage systems [38]) and multiprobe instruments, it appears that one or another solution to these problems could be found in the near future.

A greater range of angular flexibility could be achieved by placing nanotubes at an angle to each other instead of the parallel bundle alignment discussed above. In fact, such a setup could remove the need for long floppy alkane chains to link small molecules (figure 4a), or, at least, allow one to use shorter and simpler linkers (figure 4b). This flexibility, however, would have to come at a substantial additional expense: first, much more precise control of individual nanotubes would be required; second, the resulting pyramidal shape would increase the effective volume of the system and cause additional steric hindrance, which may be critical in certain cases.

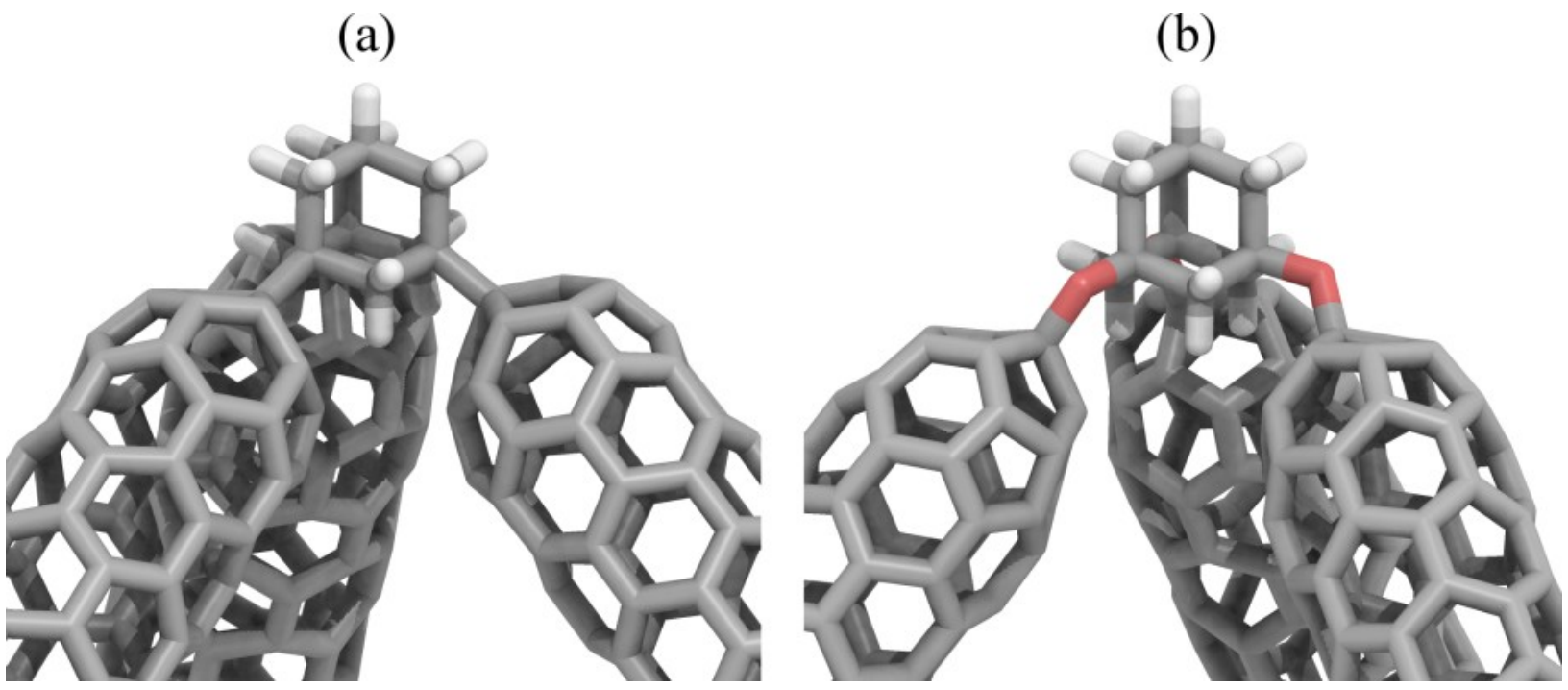

**Figure 4.** An adamantane molecule suspended on three converging nanotubes (a) directly and (b) via simple ester (C-O-C) bridges.

It should also be noted that the 4-nanotube design shown in figure 2b has one additional degree of freedom that corresponds to movement of outer and inner nanotube pairs in opposing directions. This corresponds to stretching or compression of the grafted molecule. In the present case, this effect could possibly be utilized to additionally enhance the reliability of dimer transfer either from the tool or onto the tool (when recharging), although the high stiffness of the tip molecule would probably preclude any substantial effect. More sophisticated designs could put this additional degree of freedom to use in mechanically controlled chemical reactions, or in even finer mechanical manipulation of individual molecules.

As an alternative to carbon nanotubes, other atomically-precise structural elements could be used. Recent examples include silica (0.3-0.4 nm) [39] and titania (0.4-0.5 nm) [40] nanowires. Such structures may even possess certain advantages over carbon nanotubes, such as piezo- or ferroelectricity, as well as there being fewer competing sites of possible covalent functionalization, making the assembly of complex architectures easier. Finally, besides stiff nanotubes and nanowires, more flexible chainlike structures might be utilized in future designs to build bendable manipulators. Fullerene–carbyne composite chains are one possible example [41].

## 4. Conclusions

The present communication describes a class of nanoscale parallel manipulators based on carbon nanotube bundles. The manipulators offer precise control over the position and orientation of individual molecules, thanks to the well-defined structure of constituent nanotubes and to the two additional degrees of freedom that such systems provide, compared to regular scanning probes. An important step is the choice of carbon nanotube type so as to achieve tip functionalization at predictable atomic sites. Functional molecules can then be attached by either strong covalent C-C bonds or reversible dative bonds between substitutional B and N atoms in the parts of the assembly. The designs have been demonstrated to be thermodynamically feasible, and pathways that might eventually lead to their practical implementation have been suggested. In particular, techniques to extract ultrathin carbon nanotubes from zeolite pores, or some alternative methods of free-standing ultrathin nanotube synthesis, would be desirable.

Although manipulators such as those described above can be expected to substantially improve the spatial resolution of scanning probe microscopy, the true diversity of potential applications comes from the various kinds of functional molecules that they can support. Even without the possibility to actuate individual nanotubes, rigid locking of the molecules in place will enable improved control over their position and orientation, making this approach far superior to single-nanotube imaging and manipulation [42] in terms of both versatility and precision. Here, designs that can support all 9 tooltips from the minimal toolset for positionally controlled diamond mechanosynthesis [17] have been provided. If built, they may serve as stepping stones from current scanning probe technology towards more efficient autonomous positioning systems [43] required for high-throughput deterministic manipulation of matter at the atomic scale, ultimately leading to the much anticipated prospects of machine-phase diamond [44] and graphitic [45] nanotechnology. Although the research into application of carbon nanotubes in scanning probe technologies appears to have slowed down due to practical difficulties, hopefully, the benefits from the present proposal can outweigh these and trigger further attempts to advance the needed prerequisite techniques, or stimulate the exploration of other possible ways to produce the proposed tools, possibly including some of the alternatives suggested in this communication.

## Acknowledgments

The structures were designed with the NanoEngineer-1 package (http://www.nanoengineer-1.com/). MM2 calculations were performed using the TINKER package (http://dasher.wustl.edu/tinker/). Density functional calculations were performed with the Priroda program [46]. The visualization was done using QuteMol (http://qutemol.sourceforge.net/) [47]. The author thanks D. A. Medvedev, L. A. Chernozatonskii and I. V. Artyuhov for inspiring discussions and useful suggestions on the proposal and the manuscript.